# 4*f* electron temperature driven ultrafast electron localization

Kohei Yamagami,[1,※1] Hiroki Ueda,[2] Urs Staub,[3] Yujun Zhang,[4,※2] Kohei Yamamoto,[5,※3] Sang Han Park,[6] Soonnam Kwon,[6] Akihiro Mitsuda,[7] Hirofumi Wada,[7] Takayuki Uozumi,[8] Kojiro Mimura,[8] and Hiroki Wadati[4,9]

[1]*Institute for Solid State Physics, The University of Tokyo, 5-1-5 Kashiwanoha, Kashiwa, Chiba 277-8581, Japan*
[2]*SwissFEL, Paul Scherrer Institut, 5232 Villigen PSI, Switzerland*
[3]*Swiss Light Source, Paul Scherrer Institut, 5232 Villigen PSI, Switzerland*
[4]*Graduate School of Material Science, University of Hyogo, 3-2-1 Kouto, Kamigori-cho, Ako-gun, Hyogo 678-1297, Japan*
[5]*Institute for Molecular Science, Myodaiji, Okazaki, Aichi 444-8585, Japan.*
[6]*PAL-XFEL, Pohang Accelerator Laboratory, 77 Cheongam-Ro, Nam-Gu, Pohang, Gyeongbuk 37673, Republic of Korea*
[7]*Department of Physics, Kyushu University, Motooka 744, Nishi-ku Fukuoka 819-0395, Japan*
[8]*Graduate School of Engineering, Osaka Metropolitan University, 1-1 Gakuen-cho, Nakaku, Sakai, Osaka 599-8531, Japan*
[9]*Institute of Laser Engineering, Osaka University, 2-6 Yamadaoka, Suita, Osaka 565-0871, Japan*

[※1]*Present address: Japan Synchrotron Radiation Research Institute, 1-1-1, Sayo-cho, Sayo-gun, Hyogo 679–5198, Japan*
[※2]*Present address: Institute of High Energy Physics, Chinese Academy of Sciences, Yuquan Road 19B, Shijingshan District, Beijing, 100049, China*
[※3]*Present address: National Institutes for Quantum Science and Technology, 6-6-11-901, Aramaki(others), Sendai Aoba-ku, Miyagi, 980-8579, Japan*

**Abstract:**

   Valence transitions in strongly correlated electron systems are caused by orbital hybridization and Coulomb interactions between localized and delocalized electrons. The transition can be triggered by changes in the electronic structure and is sensitive to temperature variations, applications of magnetic fields, and physical or chemical pressure. Launching the transition by photoelectric fields can directly excite the electronic states and thus provides an ideal platform to study the correlation among electrons on ultrafast timescales. The EuNi$_2$(Si$_{0.21}$Ge$_{0.79}$)$_2$ mixed-valence metal is an ideal material to investigate the valence transition of the Eu ions via the amplified orbital hybridization by the photoelectric field on sub-picosecond timescales. A direct view on the 4*f* electron occupancy of the Eu ions is required to understand the microscopic origin of the transition. Here we probe the 4*f* electron states of EuNi$_2$(Si$_{0.21}$Ge$_{0.79}$)$_2$ at the sub-ps timescale after photoexcitation by X-ray absorption spectroscopy across the Eu $M_5$-absorption edge. The observed spectral changes due to the excitation indicate a population change of total angular momentum multiplet states $J$ = 0, 1, 2, and 3 of Eu$^{3+}$, and the Eu$^{2+}$ $J$ = 7/2 multiplet state caused by an increase in 4*f* electron temperature that results in a 4$f$ localization process. This electronic temperature increases combined with fluence-dependent screening accounts for the strongly non-linear effective valence change. The data allow us to extract a time-dependent determination of an effective temperature of the 4$f$ shell, which is also of great relevance in the understanding of metallic systems' properties, such as the ultrafast demagnetization of ferromagnetic rare-earth intermetallic and their all-optical magnetization switching. In addition, our results elucidate the energetics of charge fluctuations in valence-mixed electronic systems, which provide enriched knowledge regarding the role of valence transitions and orbital hybridization for quantum critical phenomena.

**I. Introduction**

Valence transitions of 4$f$ electrons and their fluctuations are representative properties of strongly correlated 4$f$ electron systems, comparable to magnetic ordering, unconventional superconductivity, Kondo effect, and non-Fermi liquid behaviors [1]. However, a detailed description of the underlying processes, even for prototypical systems such as the element Ce, which exhibits a first-order valence transition as a function of pressure [2], remains a challenge. Many valence-transition and -fluctuating materials have fractionally occupied 4$f$ states located near the Fermi level ($E_F$) [1]. Therefore, the hybridization and the Coulomb interaction between localized 4$f$ electrons and delocalized 6$s$/5$d$ electrons play a major role in realizing the valence transition and its associated fluctuations.

Here, we focus on the Eu intermetallic compound EuNi$_2$(Si$_{0.21}$Ge$_{0.79}$)$_2$ to investigate the mechanism of its valence transition. The valence transition between Eu$^{2+}$ (4$f^7$) and Eu$^{3+}$ (4$f^6$) in tetragonal ThCr$_2$Si$_2$-type Eu compounds has been reported as a function of temperature [3-8], magnetic field [9-12], and physical and chemical pressure [3, 4, 7, 8, 13-16]. Among them, EuNi$_2$(Si$_{1-x}$Ge$_x$)$_2$ series show the temperature-driven valence transition in the range 0.5 $\leqq$ $x$ $\leqq$ 0.82 of Ge substitution [3]. The phase transition temperature ($T_v$) reaches the maximal value of ~95 K at around $x$ = 0.79, accompanied by the Kondo volume collapse [4]. An interesting aspect is the decreasing Eu valence with increasing temperature, analogous to Ce-based compounds [1,17], in contrast to the increasing lanthanide valence reported for most of the 4$f$ systems showing a valence transition, e.g., Yb-, and Sm-based compounds [18-23]. The divalent (trivalent) state of the lanthanide valence is localized (delocalized) character due to the electronic repulsion potential of the ion, which results in the large (small) ionic radius. Therefore, the valence transition in EuNi$_2$(Si$_{0.21}$Ge$_{0.79}$)$_2$ has been interpreted as a reduction in the hybridization of 4$f$ and 6$s$/5$d$ electrons at high temperatures, giving the beneficial opportunity to understand how an electron is localized from the conduction band to the localized 4$f$ band.

X-ray absorption spectroscopy (XAS) is a powerful method to investigate the equilibrium electronic state intertwined with the valence transition of lanthanide ions [24]. In addition, by combining it with optical laser excitations, the time-resolved XAS (tr-XAS) can probe nonequilibrium electronic states, currently investigated at x-ray free electron lasers (XFEL's) [25-29]. Tr-XAS allows us to track hot electrons after photoexcitation that are related to the valence transition. **Figure 1(a)** shows the schematic of the 6$s$/5$d$ conduction band and the 4$f$ band upon optical laser excitation. Since $E_F$ lies within the 4$f$ bands, localized 4$f$ electrons can be transferred to/from the conduction band after photoexcitation. Previous tr-XAS at the Eu 3$d_{5/2}$ $\rightarrow$ 4$f$ absorption edge ($M_5$ edge) for EuNi$_2$(Si$_{0.21}$Ge$_{0.79}$)$_2$ observed the photon-induced valence transition and dynamics of both Eu$^{2+}$ and Eu$^{3+}$ ions that were limited by the experimental time resolution of 70 ps [30], which is insufficient to resolve the microscopic pathways of the transition.

Recently, a fs x-ray experiment employing radiation from a hard XFEL at the Eu $2p_{3/2} \rightarrow 5d$ absorption-edge ($L_3$ edge) reported a sub-ps electron localization process, which is faster than the thermal expansion of the unit cell occurring within a few picoseconds [31]. The origin of the electron localization has been suggested to be caused by a reduction of the 4*f* hybridization with the conduction bands driven by pure electronic effects, such as screening. However, a clear understanding of the 4*f* electron dynamics at sub-ps timescale remains missing.

Lanthanides containing intermetallics are also of great interest in the field of ultrafast all optical magnetization switching (AOS). In that field, fs pulses of optical excitation can lead to a reversal of the magnetization in e.g. GdFeCo [32,33]. The excitation is expected to differently affect the magnetization 4*f* and the 3*d* metal ions, [34] and it is therefore of more general importance to understand the excited 4*f* electron system and its interaction with 3*d* transition metal ions in intermetallics. A direct view of the ultrafast behavior of 4*f* electrons by probing with fs soft X-ray pulses is therefore of great importance for a better understanding of both, AOS and ultrafast electron localization including valence transitions. In this study, we probe the sub-ps 4*f* electronic dynamics of EuNi$_2$(Si$_{0.21}$Ge$_{0.79}$)$_2$ using the $M_5$-edge tr-XAS which directly probes the 4*f* states and can show how one can directly extract the electronic 4*f* temperature which is of great importance for the ultrafast electron localization as well as basic microscopic quantity in ultrafast magnetization dynamics of lanthanide intermetallic.

**II. Experimental conditions**

Polycrystalline EuNi$_2$(Si$_{0.21}$Ge$_{0.79}$)$_2$ ingots were grown by the argon-arc melting method [9]. Magnetometry data show $T_v$ to be at ~93 K (see **Fig. S1**) [31]. The sample was mechanically cleaved in-situ in an ultrahigh vacuum at 150 K to prevent the surface layers from being oxidized.

The tr-XAS measurements were performed at the SSS beamline at the PAL-XFEL in South Korea [35-37]. An 800 nm Ti-sapphire pump laser with a pulse duration of 50 fs in full width of half maximum (FWHM), a repetition rate of 30 Hz, and a focal spot size of 600×600 μm$^2$ was used to excite the sample. The XFEL beam had a photon energy of ~1130 eV (Eu $M_5$ edge), a pulse duration of 100 fs in FWHM, a focal spot size of 80×80 μm$^2$, and a repetition rate of 30 Hz. The experimental setup is shown in **Fig. 1(b)**. The incidence angle of the XFEL pulse was 30°, and the pump laser was offset by 1° to the X rays, passing through a hole of the last optical mirror before the sample chamber. XAS data with an energy resolution of 0.7 eV were obtained by collecting the emitted electrons with a microchannel plate (MCP), placed in the horizontal plane and at 135° to the XFEL incident vector. Delay-time ($\Delta t$) scans were taken from −5.0 ps to 7.0 ps accumulating, at least, 300 shots per step for selected photon energies and each fluence. The data were normalized shot-by-shot by the incident photon flux ($I_0$) obtained from a Pt(5 nm)/Si$_3$N$_4$(200 nm) film installed upstream of the beamline.

XAS measurements were carried out at 80 K and 300 K, temperatures that are clearly below and above $T_v$, respectively. While the probing depth of the electron yield method is ~4 nm [38-40], the excitation depth of the 800 nm laser is about 25 nm for EuNi$_2$(Si$_{0.21}$Ge$_{0.79}$)$_2$, which is obtained from optical reflectivity measurements [30, 41].

**III. Eu valence changes in equilibrium and non-equilibrium.**

Characterizing the Eu $M_5$-edge XAS in equilibrium for temperatures below (80 K) and above $T_v$ (300 K) is a basis for the understanding of the nonequilibrium electronic states [see **Fig. 1c**]. The experimental data are compared with the calculated spectra from multiplet calculations for the ground states of $Eu^{2+}$ and $Eu^{3+}$ free ions [5, 30]. In the final state of the $M_5$-edge XAS, the interaction between the generated $3d$ core-hole and the $4f$ electrons with different $4f$ occupation numbers causes a 2.5 eV energy difference between the $Eu^{2+}$ (E1 = 1128.5 eV) and $Eu^{3+}$ (E2 = 1131 eV) main absorption peaks. Therefore, the XAS spectra at 300 K and 80 K are dominated by the $Eu^{2+}$ and $Eu^{3+}$ states, respectively. From the comparison to the calculations, an evaluated effective $4f$ valence ($v_{4f}$) of 2.36 ± 0.02 at 300 K and 2.71 ± 0.02 at 80 K is obtained, which is in good agreement with the previously reported values [5, 6].

**Figure 2(a)** shows $M_5$-edge tr-XAS for the lowest (0.12 mJ/cm$^2$) and highest (5.0 mJ/cm$^2$) fluences at $\Delta t$ = 0.5 ps directly probing the excited Eu $4f$ states. The unperturbed XAS data ($\Delta t$ = −5.0 ps) are also shown for comparison. The determination of $v_{4f}$ before $t_0$ results in 2.71 ± 0.02 for both fluences, in perfect agreement with the static measurements confirming the absence of any deterioration of the sample during pump-probe measurements. The photon-induced change in the XAS at 0.5 ps shown in **Fig. 2(b)** is qualitatively different between low and high fluence. There are some sign changes in the pump effect indicating larger and smaller intensities for the XAS peaks attributed to the $Eu^{3+}$ and $Eu^{2+}$, respectively. These observed pump-probe differences imply a fluence-dependent change in $v_{4f}$ of Eu ions at early times ($\Delta t$ = 0.5 ps) after the excitation. In addition, there are more complex spectral changes at early times. A simple spectral analysis from the individual divalent and trivalent spectra as done in Ref. [30] results in $v_{4f}$ of 2.67 ± 0.02 for 0.12 mJ/cm$^2$ and 2.77 ± 0.02 for 5.0 mJ/cm$^2$, suggesting the increase and decrease of the $Eu^{2+}$ population after photoexcitation at low and high fluence, respectively. This is quantitatively consistent with the previous report based on $L_3$-edge tr-XAS [31].

By adjusting the photon energy to the $Eu^{2+}$ peak (E1), we directly investigated the time evolution of XAS spectral intensity changes ($\Delta I$) [see **Fig. 2(c)**]. A stepwise increasing component of the $Eu^{2+}$ peak was observed from the low fluence region (0.12-0.35 mJ/cm$^2$). Furthermore, a prominent decrease in the $Eu^{2+}$ signal is captured at earlier times through the high fluence region (3.75-5.0 mJ/cm$^2$). This fluence dependence of the time evolution is analogous to that reported in $L_3$-edge tr-XAS [31]. This continuous change in the time evolution as a function of laser fluences can be described by a phenomenological model with different exponential components given by

$$\Delta I(\Delta t) = G(\Delta t) \otimes [(I_1 + I_2)H(\Delta t)] \quad \ldots \quad (1)$$
$$I_i = A_i\left(1 - e^{-\Delta t/\tau_i}\right) \quad \ldots \quad (2)$$

where $A_i$ and $\tau_i$ ($i$ = 1, 2) denote an amplitude and a time relaxation, respectively. $G(\Delta t)$ is a Gaussian function reflecting the time resolution and $H(\Delta t)$ is the Heaviside function [see **Fig. S2**]. The first component $I_1$ represents a slow process, possibly the response for the energy transfer from the excited electronic system to the lattice. The second component $I_2$ should then be associated with the initial charge delocalization/localization process. The obtained $A_i$ and $\tau_i$ parameters are plotted as a function of excitation fluence in **Figs. 2(d) and 2(e)**. The amplitude of the first component shows a linear laser fluence dependence with a relaxation time being almost constant, which is consistent with the previous findings of 70 ps time resolution tr-XAS [30]. The amplitude of the second component is strongly non-linear with a fluence-dependent increase in the relaxation time and flipping its sign, consistent with the previous findings at the $L_3$ edge tr-XAS [31]. This directly reveals that the photoexcited $4f$ electronic character changes delocalized nature to localized nature with increasing laser fluence at the sub-picosecond timescale. Note that the time evolution of XAS spectral intensity changes at the photon energy of the Eu$^{3+}$ peak (E2) can roughly be reproduced by flipping the sign of $\Delta I(\Delta t)$ in Eq.(1), indicating that 4$f$ electrons at both Eu$^{2+}$ and Eu$^{3+}$ ions are simultaneously photoexcited [see **Fig. S3**].

**IV. Nonequilibrium 4$f$ electronic structure analyzed by the interconfigurational fluctuation model.**

A significant advantage of $M_5$-edge tr-XAS is its direct sensitivity to Eu 4$f$ electronic states. Upon increasing laser fluence at the early delay times, the Eu$^{3+}$ main peak at E2 decays and another spectral feature at E3 = 1132.5 eV develops, in addition to the sign change of the Eu$^{2+}$ peak at E1 as shown in **Fig. 2(b)**. It clearly shows that there are competing non-linear processes such as e.g., charge localization/delocalization that differently depend on the excitation fluence. Note that these photoinduced processes are sufficiently faster than the volume expansion (~3 ps) observed previously [31]. One plausible mechanism is that the electronic temperature increase of the 4$f$ states due to the excitation, which directly leads to an increase in the Eu$^{2+}$ state population. The other possible mechanism is related to the increase of the electronic temperature of the conduction band, which leads to screening that will differently affect the Eu$^{2+}$ and Eu$^{3+}$ states. These mechanisms can compete or coexist, and possibly result in charge localization/delocalization depending on laser fluence, as discussed later.

To gain more insight into these scenarios, we consider the excited $J$ multiplets of the Eu$^{3+}$ ions within the free-ion approximation that allows us to evaluate the population of the excited states. A multiplet calculation for free Eu$^{3+}$ ions with the different occupations of higher-lying $J$ multiplets predicts that the $M_5$-edge XAS main peak becomes broader [**Fig. 3(a)**], and thus the population can be evaluated from the spectral shape. The electronic 4$f$ temperature can be addressed through the interconfigurational fluctuation (ICF) model, known as a high-temperature model for incoherent thermal charge fluctuations used for rare-earth metals [9, 13-15, 42-44]. The energy level scheme is shown in **Fig. 3(b)**. $E_J$ is the energy of the of Eu$^{3+}$ atomic multiplet states with the total angular momentum $J$ neglecting their crystal field splitting [**Fig. 3(b)**]. $E_{ex}$ is the energy to take one electron out of the conduction band and to put it into the 4$f$ levels converting the Eu valence state from 3+ to 2+. In this model, the populations of the Eu$^{2+}$ and Eu$^{3+}$ states, represented as $p_2$ and $p_3$, respectively, are given by the Boltzmann statistics

$$\frac{p_2}{p_3} = \frac{N_J \exp\left(-\frac{E_{ex}}{k_B T^*}\right)\Big|_{J=7/2}}{\sum_{J=0}^{3} N_J \exp\left(-\frac{E_J}{k_B T^*}\right)} = \frac{8 \exp\left(-\frac{E_{ex}}{k_B T^*}\right)}{1 + 3 \exp\left(-\frac{39.6}{k_B T^*}\right) + 5 \exp\left(-\frac{118.9}{k_B T^*}\right) + 7 \exp\left(-\frac{237.8}{k_B T^*}\right)} \quad \cdots \quad (3),$$

where $k_B$ is the Boltzmann constant. The degeneracy $N_J = 2J + 1$ is adopted as the coefficient of each Boltzmann distribution for $J = 7/2$ state of Eu$^{2+}$ and $J = 0, 1, 2$, and 3 states of Eu$^{3+}$ [**Fig. 3(a)**] and $p_2 + p_3 = 1$. $T^*$ (= $\sqrt{T^2 + T_{4f}^2}$) is the effective temperature including the temperature in the system ($T$) and the constant value of the broadening of the 4$f$ state ($T_{4f}$). Here, $E_{ex}$ and $T^*$ are the tuning parameters to reproduce the XAS data and $v_{4f}$ is given by $v_{4f} = 3 - p_2$. The excited Eu$^{3+}$ spectra for a given electronic temperature $T^*$ were obtained by summing the XAS spectra of each $J$ state, shown in **Fig. 3(a)**, multiplied by their population. **Figure 3(c)** shows the experimental $M_5$-edge tr-XAS for 0.12 mJ/cm$^2$ at 80 K and $\Delta t$ = 0.5 ps together with equilibrium data for comparison

($\Delta t$ = –5.0 ps) as well as the calculated tr-XAS based on the ICF model. Spectral differences between excited and ground states for experimental and calculated data sets are shown in the bottom panel of **Fig. 3(c)**. The calculated spectra obtained by the ICF model reproduce well the experimental data. As referring from the previous $L_3$-edge XAS [9], $E_{ex}$ and $T^*$ in the equilibrium XAS data are estimated to be 48 meV and 203 K, respectively. Furthermore, it confirms that the calculated $v_{4f}$ = 2.71 matches to our static $M_5$-edge XAS result at $T$ = 80 K [**Fig. 1(c)**]. For the excited state at $\Delta t$ = 0.5 ps, $T^*$ is estimated to be 222 K when keeping $E_{ex}$ = 48 meV.

**Figures 4(a) and 4(b)** show the experimental and calculated data of the tr-XAS and their spectral deviations from the equilibrium data for selected laser fluences at $\Delta t$ = 0.5 ps and $T$ = 80 K, respectively. The experimental data are well described by a fit to the ICF model with the following restrictions on $T^*$ and $E_{ex}$. Since the initial photo absorption process heats only the electronic structure, the temperature $T^*$ is assumed to be proportional to the laser fluence. Its slope is derived from the experimental data for the 0.12 mJ/cm² at $\Delta t$ = –5.0 ps (equilibrium) and 0.5 ps (non-equilibrium). While $E_{ex}$ is freely tuned to reproduce the experimental data. The occupation of each multiplet state plotted as a function of laser fluence in **Fig. S3(a)** indicates strongly non-linear behavior, except for the $Eu^{3+}$ $J$ = 2 state. $E_{ex}$ is found to increase approximately linear with fluence except for low fluences, as shown in **Fig. S3(b)**. Note that the increase in $E_{ex}$ at early times ($\Delta t$ = 0.5 ps) is not an absolute energy increase of the $Eu^{2+}$ states but is in respect to the changes of the $Eu^{3+}$ states population. The non-linear effect on the valence change, as obtained in the time traces at 0.5 ps, is a consequence of $E_{ex}$ being constant at small fluences and the intrinsic non-linear behavior of the population in the Boltzmann statistics.

To better visualize the origin of the sign change in the electron localization/delocalization as a function of fluence at 0.5 ps, we plot the population changes of the $Eu^{2+}$ and $Eu^{3+}$ while keeping $E_{ex}$ fixed in **Fig. 4(c)**. This allows us to qualitatively understand the time traces taken at the main XAS peak of the $Eu^{2+}$ (E1). **Figure 4(d)** shows the $T^*$ − $E_{ex}$ mapping of $v_{4f}$. At low fluence, the increase in $T^*$ just leads to a population increase of the divalent state. For high fluence, the ultrafast increase in $T^*$ can lead to a increase of the valence state; however, as the electronic temperature cools due to energy transfer to the lattice, it leads again to reduced $E_{ex}$ and the population of the $Eu^{2+}$ increases again.

## IV. Discussions: Ultrafast 4*f* localization and related magnetic phenomena

These findings allow us to draw a semi-quantitative picture on the valence change and its mechanism of the photoinduced valence transition. After the initial excitation of the electronic structure, the energy transfers in approximately 3 ps to the optical phonons and later to acoustic modes, as found previously by ultrafast X-ray diffraction [31]. The initial excitation does not only inject energy into the different 4*f* states but also increases the electron mobility (electron temperature of the conduction band) leading to screening effects, which is responsible for a change of $E_{ex}$. Attosecond experiments have previously shown that screening leads to an electron localization in Cu metal [45]. In our case, however, the turnover from charge localization to delocalization for increasing fluence is intrinsically contained in the population changes by solely increasing the effective temperature in the ICF model alone. On the other hand, the ICF model is rather limited, as the 4*f* bandwidth, which contains the important hybridization between conduction electrons and localized 4*f* states, is only described by its contribution to the effective temperature. A quantitative understanding of the intrinsic effect of screening and effective 4*f* temperature for valence transition requires a description that includes hybridization and interaction of conduction electrons with localized 4$f$ and delocalized 5$d$ Eu states more adequately, which is not possible in the phenomenological ICF model used here. However, it shows that the spectral shape changes allow a quantitative determination of the 4$f$ electron temperature. In general, this is of great interest in studies on ultrafast magnetization dynamics, in particular for magnetic 4$f$ materials such as those showing all-optical magnetization switching, e.g., GdFeCo [32,33,46-48]. A determination of the electronic temperature of the 4$f$ electron systems in 4$f$ intermetallics materials or multilayers will be of great importance understand the demagnetization processes as well as inter-site spin processes [49] believed to be relevant for ultrafast magnetization dynamics. In addition, it has been recently shown that excited 4f multiplet (orbital) states are important for the magnetization dynamics of Tb metal [50].

## V. Conclusion

The time-resolved Eu $M_5$-edge XAS quantifies the valence change induced by an optical excitation in a valence fluctuating EuNi$_2$(Ge/Si)$_2$ intermetallic. The approach of direct detection of the 4$f$ states through spectral changes allows addressing the effective electronic temperature of the 4$f$ states through the interconfigurational fluctuation model. The obtained results support the view that an ultrafast temperature increase of the 4$f$ system is responsible for initial charge localization in the 4$f$ states at low fluences, which brings a novel spin on the physics of 4$f$ valence fluctuations. For higher fluences, screening effects must be considered. Beyond valence fluctuation/transition physics, the determination of the 4$f$ electron temperature by means of the XAS method will also be of significant interest to mixed 4$f$ - 3$d$ systems that show ultrafast demagnetization or all-optical magnetization switching.

**Acknowledgements:**

We thank H. Watanabe and Y. Yokoyama for the fruitful discussions. The experiments were performed at SSS end-station of PAL-XFEL (proposal no. 2019-2nd-SSS-014) funded by the Ministry of Science and ICT of Korea. This work was supported by MEXT Quantum Leap Flagship Program (MEXT Q-LEAP) Grant Number JPMXS011806868. H.U. was supported by the National Centers of Competence in Research in Molecular Ultrafast Science and Technology (NCCR MUST-No. 51NF40-183615) from the Swiss National Science Foundation and from the European Union's Horizon 2020 research and innovation program under the Marie Skłodowska-Curie Grant Agreement No. 801459 – FP-RESOMUS.

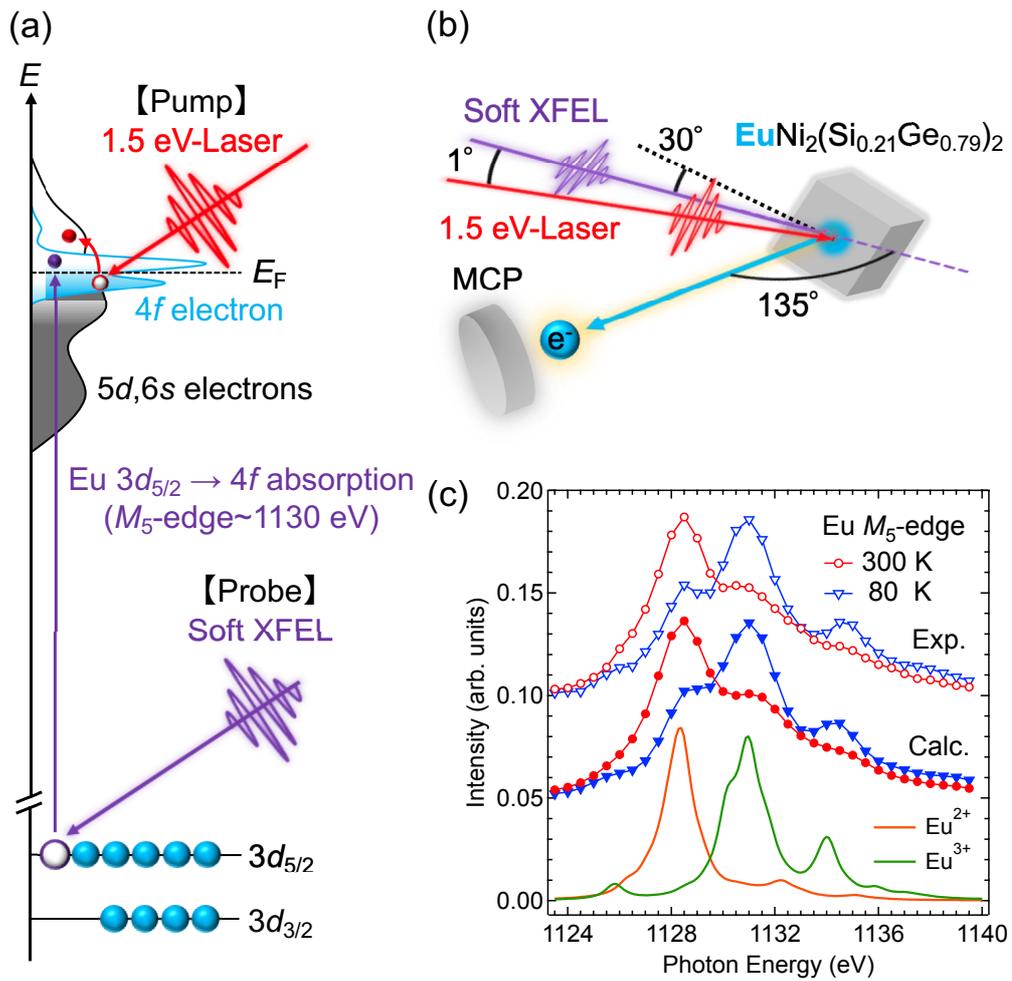

FIG. 1. (a) Schematic of the electronic bands including a core hole of the XAS probe, modifications of the $4f$ states upon laser excitation (red arrows), and $3d_{5/2} \rightarrow 4f$ core-level absorption ($M_5$-edge) process (purple arrows). (b) The experimental setup of the tr-XAS. (c) XAS around the Eu $M_5$ edge at 300 K and 80 K (upper spectra) crossing the mixed-valence transition temperature. The middle spectra are the atomic multiplet calculations broadened to match experimental data. The calculated spectral components of the $Eu^{2+}$ and $Eu^{3+}$ free ion spectra are shown at the bottom.

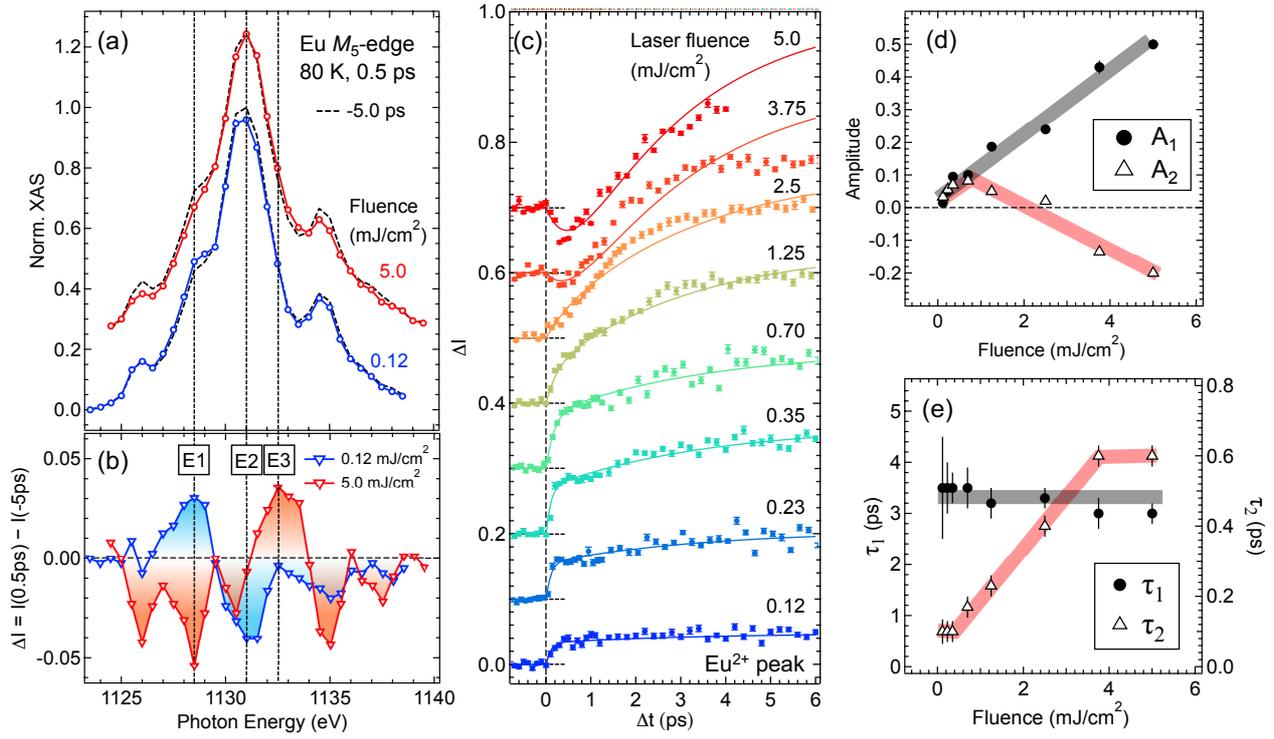

FIG. 2. (a) The Eu $M_5$-edge tr-XAS spectra at 80 K under low (0.12 mJ/cm$^2$, blue) and high fluences (5.0 mJ/cm$^2$, red) with and without (black, taken at $\Delta t = -5$ ps) laser excitation at early times ($\Delta t = 0.5$ ps). (b) XAS spectral intensity changes ($\Delta I$) at 0.5 ps. In panel (a) and (b), the vertical dashed lines denote the photon energies, labeled as E1 = 1128.5 eV, E2 = 1131 eV, and E3 = 1131.5 eV, which are discussed in the main text. (c) Ultrafast $\Delta I$ change as a function of $\Delta t$ for selected fluences at 80 K of the Eu$^{2+}$ peak (E1). The colored solid lines are best fits to two exponential functions. (d), (e) Laser fluence dependence of the amplitude $A_i$ and the recovery time $\tau_i$ ($i = 1, 2$). The solid lines guide to the eye.

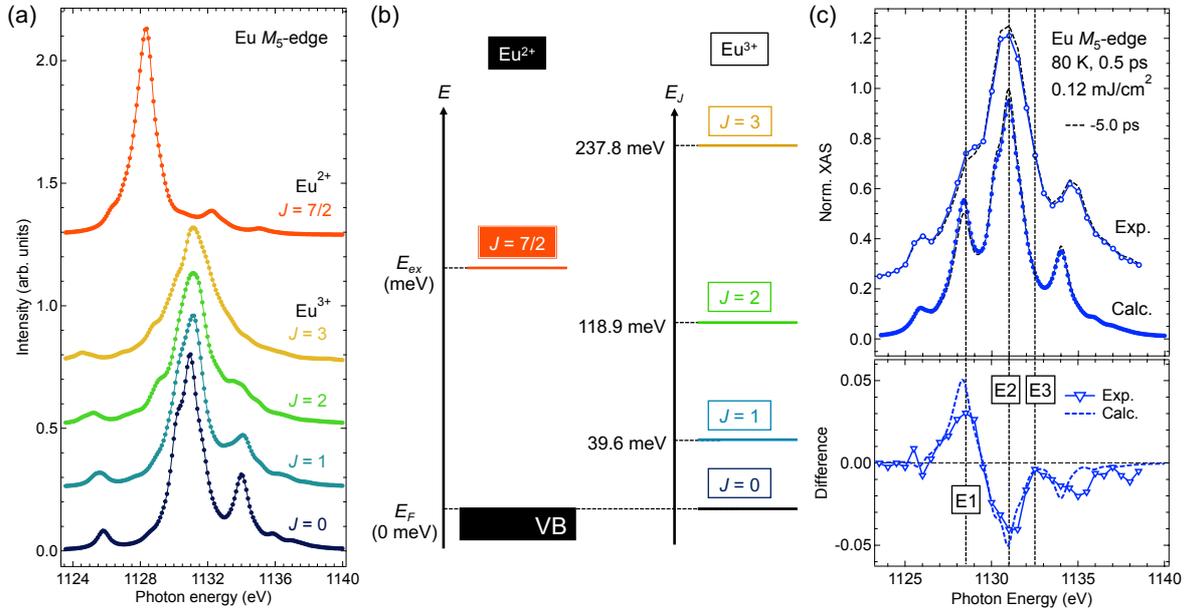

FIG. 3 (a) The atomic multiplet calculations of each $J$ states. (b) Energy level of the ground and excited states for $Eu^{2+}$ and $Eu^{3+}$ free ions. $E_{ex}$ is the energy of $Eu^{2+}$ $J = 7/2$ state from the Fermi level, which is at the $Eu^{3+}$ $J = 0$ state. (c) The experimental and calculated Eu $M_5$-edge XAS spectra at 80 K, 0.12 mJ/cm$^2$, and $\Delta t = 0.5$ ps. Black dashed curves indicate data without laser excitation taken at $\Delta t = -5.0$ ps (equilibrium). The spectral difference from the equilibrium data is shown in the bottom panel.

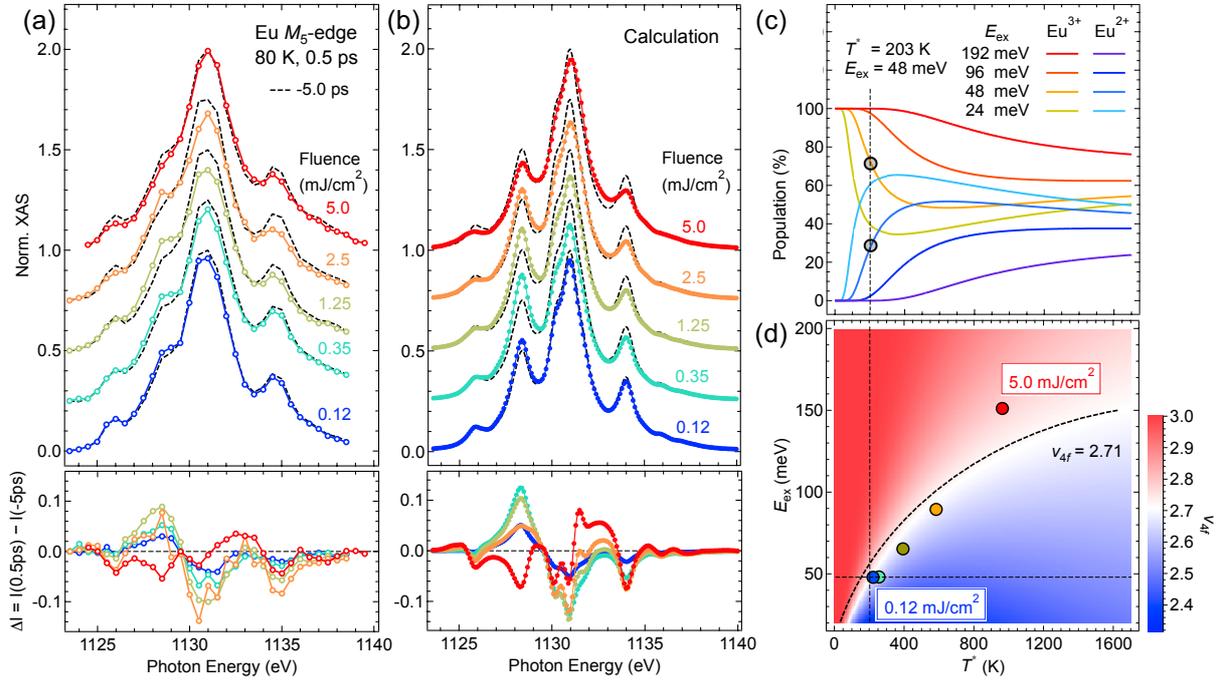

FIG. 4 (a) The $M_5$-edge XAS and $\Delta I$ spectra for selected fluences at 80 K and $\Delta t = 0.5$ ps, together with equilibrium data taken at $\Delta t = -5.0$ ps. (b) The calculated XAS and their difference from equilibrium spectrum to reproduce well the experimental results. (c) Calculated $Eu^{2+}$ and $Eu^{3+}$ populations at different $E_{ex}$ as a function of $T^*$. The vertical black dashed line with circle marks approximately the equilibrium case (80 K). (d) The $T^* - E_{ex}$ mapping of the Eu valence. The circles denote the obtained values for each fluence from spectral fitting based on the ICF model. The curved dashed line corresponds to the Eu valence of 2.71. Horizontal line cuts at fixed $E_{ex}$ = 24 meV, 48 meV, 96 meV, 192 meV, denoting non-linear Eu valence change as a function of $T^*$, are shown in **Fig. S5**.

# Supplemental information of

# 4*f* electron temperature driven ultrafast electron localization

Kohei Yamagami,[1,※1] Hiroki Ueda,[2] Urs Staub,[3] Yujun Zhang,[4,※2] Kohei Yamamoto,[5※3] Sang Han Park,[6] Soonnam Kwon,[6] Akihiro Mitsuda,[7] Hirofumi Wada,[7] Takayuki Uozumi,[8] Kojiro Mimura,[8] and Hiroki Wadati[4,9]

[1]*Institute for Solid State Physics, The University of Tokyo, 5-1-5 Kashiwanoha, Kashiwa, Chiba 277-8581, Japan*
[2]*SwissFEL, Paul Scherrer Institut, 5232 Villigen PSI, Switzerland*
[3]*Swiss Light Source, Paul Scherrer Institut, 5232 Villigen PSI, Switzerland*
[4]*Graduate School of Material Science, University of Hyogo, 3-2-1 Kouto, Kamigori-cho, Ako-gun, Hyogo 678-1297, Japan*
[5]*Institute for Molecular Science, Myodaiji, Okazaki, Aichi 444-8585, Japan.*
[6]*PAL-XFEL, Pohang Accelerator Laboratory, 77 Cheongam-Ro, Nam-Gu, Pohang, Gyeongbuk 37673, Republic of Korea*
[7]*Department of Physics, Kyushu University, Motooka 744, Nishi-ku Fukuoka 819-0395, Japan*
[8]*Graduate School of Engineering, Osaka Metropolitan University, 1-1 Gakuen-cho, Nakaku, Sakai, Osaka 599-8531, Japan*
[9]*Institute of Laser Engineering, Osaka University, 2-6 Yamadaoka, Suita, Osaka 565-0871, Japan*

[※1]*Present address: Japan Synchrotron Radiation Research Institute, 1-1-1, Sayo-cho, Sayo-gun, Hyogo 679–5198, Japan*
[※2]*Present address: Institute of High Energy Physics, Chinese Academy of Sciences, Yuquan Road 19B, Shijingshan District, Beijing, 100049, China*
[※3]*Present address: National Institutes for Quantum Science and Technology, 6-6-11-901, Aramaki(others), Sendai Aoba-ku, Miyagi, 980-8579, Japan*

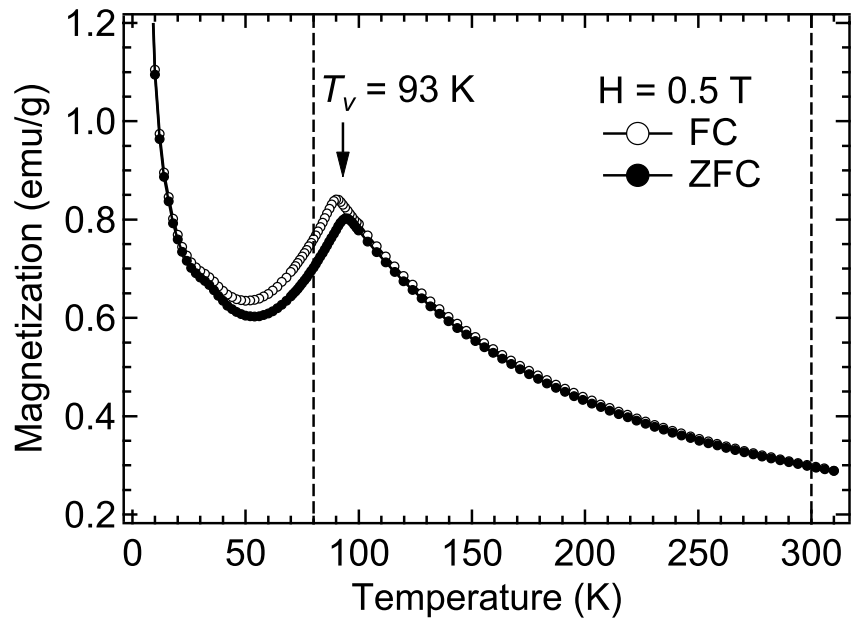

FIG. S1 The temperature dependence of magnetization of EuNi$_2$(Si$_{0.21}$Ge$_{0.79}$)$_2$. The recorded data at 0.5 T shows the transition temperature around 93 K under both field-cooling and zero-field-cooling processes. Note the tr-XAS experiments were performed without a magnetic field.

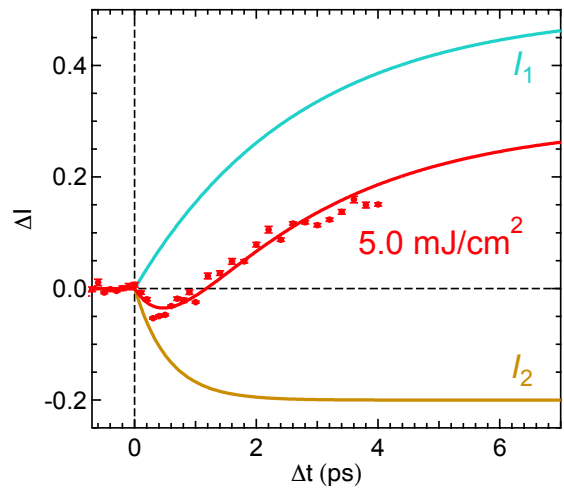

FIG. S2 The two exponential curves employed to fit the tr-XAS data at 5.0 mJ/cm$^2$.

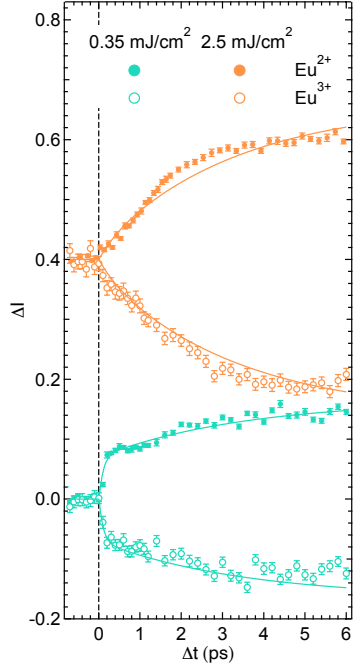

FIG. S3 The comparison with ultrafast $\Delta I$ change as a function of $\Delta t$ for selected fluences of 0.35 mJ/cm$^2$ and 2.5 mJ/cm$^2$ at the photon energy of the Eu$^{2+}$ peak (E1) and Eu$^{3+}$ peak (E2). The colored solid lines are best fits to two exponential functions described as Eq.(1).

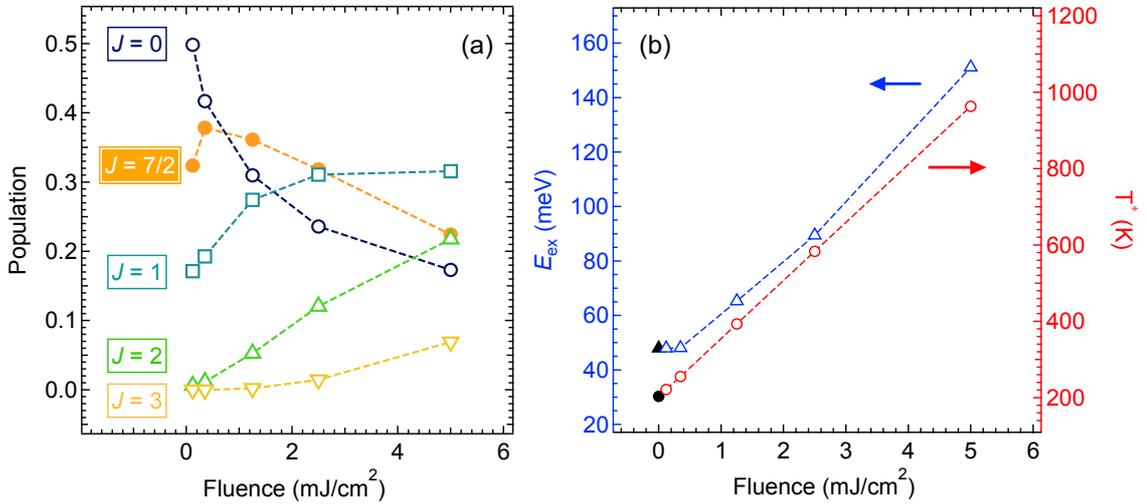

FIG. S4 (a) The fluence dependence of the occupation probabilities of each $J$ states. (b) The fluence dependence of the $E_{ex}$ (blue triangle) and $T^*$ (red circle) obtained by the ICF model. In panel (b), the data without laser excitation is also plotted as black filled makers.

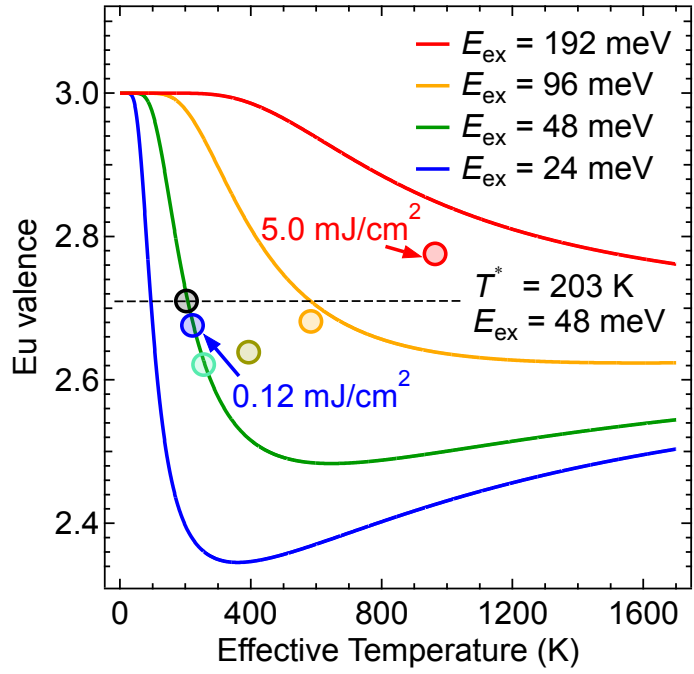

FIG. S5 Non-linear effective-temperature dependence of the Eu valence at selected $E_{ex}$. The colored marks denote the obtained values for each fluence from spectral fitting based on the ICF model.